\documentclass[prb,twocolumn]{revtex4}
\usepackage{amssymb}
\usepackage{graphicx}
\usepackage{amsmath}

\begin{document}

\title{Polarized Raman study of the phonon dynamics in Pb(Mg$_{1/3}$Nb$%
_{2/3} $)O$_{3}$ crystal}
\author{Oleksiy Svitelskiy}
\altaffiliation{On leave from the Institute of Semiconductor Physics, Kyiv 03028, Ukraine}
\affiliation{Department of Physics, Lehigh University, Bethlehem, PA 18015, USA}
\author{Jean Toulouse}
\affiliation{Department of Physics, Lehigh University, Bethlehem, PA 18015, USA}
\author{Z.-G.Ye}
\affiliation{Department of Chemistry, Simon Fraser University, Burnaby, BC, V5A 1S6,
Canada}
\pacs{}

\begin{abstract}
Pb(Mg$_{1/3}$Nb$_{2/3}$)O$_{3}$ is one of the simplest representatives of
the class of lead relaxors and often serves as a model system for more
complicated compounds. In this paper we analyse both polarized and
depolarized Raman scattering spectra, measured in the temperature range
between 1000 and 100~K, by multiple peak decomposition. Basing on this
analysis, we present the picture of transformations in the crystal and
the nature of the spectral lines. According to our model, the formation of
the Fm3m symmetry in the chemically ordered regions as well as the
appearance and freezing of the polar nanoregions are the consequences of the
same phenomenon: ion off-centered displacements and their reorientational
thermal motion. The lowering temperature progressively limits this motion
from eight to four, two and only one site. Raman scattering exists due to
the presence of the Fm3m and polar clusters. It is a product of the
multiphonon interaction process, mediated by the presence of the dynamic and
static disorder.
\end{abstract}

\maketitle

\section{Introduction}

Pb(Mg$_{1/3}$Nb$_{2/3}$)O$_{3}$ (PMN) is one of
the earliest members of the growing family of lead
ferroelectric relaxors, which, due to its application potential, has been
attracting attention of researchers for many years. However, despite 
significant efforts, many properties of lead compounds are still
unclear and the very nature of the relaxor behavior has yet to be understood. The difficulties stem from the high complexity of these materials
with a high degree of compositional, chemical and structural disorder.

Pb(Mg$_{1/3}$Nb$_{2/3}$)O$_{3}$ has a perovskite structure with, on average,
a cubic Pm3m symmetry, with Mg$^{2+}$ and Nb$^{5+}$ ions interchanging on B
sites. Measurements of the dielectric constant show that the crystal does
not undergo a sharp transition to a ferroelectric phase. Instead, the dielectric
constant exhibits a broad maximum at T$%
_{\max }\sim270$~K\cite{bokov}. Below T$_{d}\sim620$~K, it
deviates from Curie-Weiss law\cite{viehland-weiss} and, at T$\lesssim $%
350~K, exhibits strong dispersion\cite{viehland}. The structure
of PMN is not homogeneous. If Mg and Nb occupation ratio were uniformly
and ideally equal 1:2 throughout the crystal, it would have a rhombohedral
symmetry R3m. But, a network of superstructure clusters with face
centered cubic symmetry (Fm3m), destroys
the picture. The size of these clusters is of the order of 2-3 nm; the
distance between the centers of the neighboring clusters is near 2.5~nm\cite%
{boulesteix,harmer}.

The most widely accepted
hypothesis, so-called \textquotedblleft space-charge model\textquotedblright
, associates these Fm3m clusters with antiferroelectrically ordered\cite{miao} areas where Mg:Nb occupation ratio is equal to 1:1, i.e. with local composition Pb(Mg$%
_{1/2}$Nb$_{1/2}$)O$_{3}$\cite{husson-11, hilton}. Such 1:1 clusters carry an excessive negative charge, which is compensated by the positively charged host matrix. Network of these charges
acts as a source of random fields and may prevent the occurrence of the
\textquotedblleft normal\textquotedblright\ phase transition. 
Alternatively, the Fm3m clusters might be explained by the local order in the form of Pb[B$_{1/2}^{2+}$B$_{1/3}^{5+}$]$_{1/2}$[B%
$^{5+}$]$_{1/2}$O$_{3}$ \cite{akbas, davies}. This hypothesis avoids assumption of the considerable space charges and calls for other reasons to explain relaxor behavior, including the random fields due to the disorder-induced local anisotropy\cite{Blink}, effects related to different
ionic radii\cite{chen} or both\cite{egami}.

The presence of areas with special order causes crystalline structure to
change in an intricate way. At the high temperature limit, the crystal
has a uniform primitive cubic (Pm3m) structure. Lowering the temperature leads
to distinguishability between Mg$^{2+}$ and Nb$^{5+}$ sites and to formation in the ordered regions of a face centered (Fm3m) superstructure. The temperature of the Fm3m clusters formation has yet to be determined. Husson's data suggests they presence at T$\sim $%
850~K \cite{husson}. Our data (see below) raise the upper limit to a temperature higher than 1000~K.

\begin{table*}[ht]
\caption{Interpretations of Raman scattering spectrum from PMN offered by
various research groups. Note, many of the entries in the table mutually
exclude each other.}
\label{RamanConc}%
\begin{ruledtabular}
\begin{tabular}{lllllll}
Research & 45 cm$^{-1}$ & 130 cm$^{-1}$ & 260 cm$^{-1}$ & 420 cm$^{-1}$ & 500-600  & 780 cm$^{-1}$   \\
Group    &              &               &               &               & cm$^{-1}$ &            \\
\hline
\hline
Husson\cite{husson}   &\multicolumn{2}{c}{Pb-O stretching modes} & O-B-O         & Mg-O-Mg       & Nb-O...Nb  & Nb-O...Mg   \\
(1990)   &              &               & bending       & stretching    & stretching & stretching   \\
\hline
Dimza\cite{dimza}    &              &               &               &               &            & Coexistence of tetragonal   \\
(1992)   &              &               &               &               &            & and rhombohedral phases   \\
\hline
Krainik\cite{krainik} (1993)   & TO$_{1}$     &               &               &               &            &          \\
\hline
Idink\cite{idink}    & \multicolumn{6}{c}{All spectrum is due to one phonon processes, originates}  \\
(1994)   & \multicolumn{6}{c}{from rhombohedral symmetry clusters and consists of 9 modes}                 \\
\hline
Marssi\cite{marssi, marssiA, marssiB, marssiC}   & TO$_{1}$     & TO$_{2}$      & TO$_{3}$      & TO$_{4}$      &            & TO$_{2}$+TO$_{4}$ or LO$_{4}$ (1996) \\
(1996-98)     &              &               &               &               &            & Nb-O-Mg or LO$_{4}$ (1998)  \\
\hline
Jiang\cite{jiang} (1999)     &\multicolumn{3}{c}{Due to positional disorder of lead}&\multicolumn{3}{c}{Due to Fm3m clusters and vibration of oxygen octahedra} \\
\hline
        \multicolumn{7}{c}{All spectrum originates from Fm3m clusters, similarly to the case of PbSc$_{1/2}$Ta$_{1/2}$O$_{3}$}\\
         & VV-E$_{g}$?? &               &               &               &            &   \\
Lushnikov\cite{lushnikovA, lushnikovB}& VH-2T$_2g$   &               &               &               &            & A$_{1g}$  \\
Siny\cite{sinyMRS, sinyJRS, sinyF-A, sinyF-B, sinyF-Rev}     & or           & \multicolumn{4}{c}{Disorder-induced scattering from silent modes} & Breathing mode  \\
(1999)   & VV and one VH &\multicolumn{4}{c}{and second-order scattering} &of oxigen  \\
         & have fractal & & & &  & \\
         & character    & & & & & \\
         & another VH-T$_{2g}$ & & & & & \\
\hline
Dujovne\cite{Dujovne} (2002)& TA+disorder& & & & & \\
\end{tabular}
\end{ruledtabular}
\end{table*}

Cooling the sample leads to the nucleation of polar nanoregions
(PNR's) characterized by the local distortions. These distortions are
responsible for the deviation of the dielectric constant from Curie-Weiss
law at T$_{d}$\cite{viehland-weiss}. Their presence also causes a non-linearity in the temperature dependence of the optical refraction index%
\cite{burnsRefraction}. Size of the PNR's is equal to 30-80~\AA\, depending on temperature\cite{takesue}. The appearance of polar distortions is a consequence of the ion off-centering that is a common
property of many perovskites. By structural similarity, one could expect that the mechanism of the polar regions formation in PMN is analogous to the one in KTa$_{1-x}$Nb$_{x}$O$_{3}$
(KTN)\cite{toulouse}. However, unlike KTN where Nb$^{5+}$ is the only
off-centered positive ion, PMN represents a more difficult case. Here, not only 
\textit{both ions of B-type}, Mg$^{2+}$ and Nb$^{5+}$, are shifted by $\sim $%
0.1-0.2~\AA , but also Pb$^{2+}$, \textit{ion of A-type}, is off-centered by 
$\sim $0.25-0.35~\AA \cite{Bonneau, mathan, verbaere, uesu, vakhrushev}. The
B-ions tend to shift along a (111) direction, thus having eight
equivalent positions. The shifts of the Pb$^{2+}$-ions exhibit a spherically symmetric distribution\cite%
{vakhrushev}. It is often assumed\cite{husson, Blink, chen, egami} that
Nb$^{5+}$, due to its position in the cell and small radius, acts as the main ferroelectric agent.

At high temperature, the off-centered ions are free to reorient among all
allowed off-centered positions. Cooling down and the growing role of the
electric interactions causes the appearance of correlated regions
consisting of several neighboring cells. Each such region is characterized
by a giant electric dipole moment and a local distortion from the cubic
symmetry. The PNR's are capable of reorientational motion as whole units. In KTN, the
development of these regions leads, at some critical temperature,
to a phase transition. However, in PMN, the polar regions
develop under constraints imposed by the local anisotropy.
As a result, no ferroelectric phase transition, but a glasslike freezing of
the polar regions is observed\cite{Blink}. A bias electric field
of magnitude $\sim1.8$~kV/cm, applied in the (111) direction, is sufficient
to compensate for the influence of the frustrating effects and to induce, at
T$_{do}\thicksim210$~K, a transition to the rhombohedral R3m ferroelectric
phase\cite{Ye}.

Raman spectra of PMN have been measured and reported by several research
groups\cite{husson,ohwa}. But, the high complexity of the processes caused
difficulties in their interpretation, especially due to possible coexistence of clusters with Pm3m, Fm3m and R3m symmetry. It has been shown that the major spectral lines are of first-order character\cite{burns}. In cubic crystals first-order Raman scattering is prohibited by symmetry.
Explanation of the appearance of first-order scattering \textit{in principle} is the one side of the problem. Another side is the assignment of \textit{particular} lines.
The question of the soft ferroelectric mode (TO$_{1}$) and the strong Raman
line at $\sim $45~cm$^{-1}$, which is close to where the TO$_{1}$ is
expected to be found,\ has been like a stumbling block for investigators.
Interpretations of the PMN\ Raman spectrum offered by various research groups
are summarized in Table \ref{RamanConc}. This table reflects the 
variety of mutually excluding opinions ranging from those that explain the
light scattering by \textit{disorder in general} (groups of Husson\cite%
{husson}, Krainik\cite{krainik}, Marssi\cite{marssi,marssiA,marssiB,marssiC}, Dujovne\cite{Dujovne}) to those that connect it to the presence of some \textit{particular type
of disorder}, the one that can lead to the appearance of clusters with
symmetry allowing first-order scattering, like face centered cubic (Jiang%
\cite{jiang}, Lushnikov\cite{lushnikovA,lushnikovB}, Siny\cite%
{sinyMRS,sinyJRS,sinyF-A,sinyF-B,sinyF-Rev}), rhombohedral\ (Idink\cite%
{idink}) or tetragonal (Dimza\cite{dimza}). The assignement of particular lines is even more contradictory. Recently made first-principle calculations \cite{pros2} can confirm only few of them. We believe, the arguments might be resolved by taking into account the
possibility of coexistence of different mechanisms, each playing a role in
the scattering processes.

Recently published low branches of phonon dispersion curves\cite%
{GehWak,WakStock,GehVak,NaberVak}, measured by neutron scattering, must be helpful for the phonon assignment of some of the Raman lines. According to these data, the strong 45~cm$^{-1}$ line has an
energy close to the energy of the zone center (ZC) TO$_{1}$ or the zone boundary (ZB) TA
phonon. Lowering the temperature, TO$_{1}$ zone center phonon
exhibits softening, at T$\sim$T$_{d}$~K, due to the drastic
increase of damping, it disappears. The overdamping continues down to T$_{do}\sim210$~K\cite{GehWak}. It also causes significant, almost sixfold, broadening of the TA phonon
branch\cite{koo}. It seems reasonable to expect that the Raman line at 45~cm$%
^{-1} $ is also affected by the described phenomenon. However, none of
the known Raman reports show any significant effects that could
be attributed to this overdamping. 

The formation of the lower-symmetry clusters in the host crystal is
accompanied by relaxational and reorientational dynamics. This dynamics
should be reflected in the light scattering spectrum either directly or through
interaction with the phonon modes. Theoretical considerations\cite{MichelA} show that
internal relaxational motion leads to the appearance of the central peak
(CP) in the Raman spectrum. Interactions of this motion with phonons can
cause softening of them\cite{MichelA}. These predictions have been tested on
several types of crystals with internal degrees of freedom, like
alkali-halide-cyanide (K(CN)$_{x}$-KBr$_{1-x}$, K(CN)$_{x}$-KCl$_{1-x}$)\cite%
{RoweA,RoweB,LutyA,LutyB} compounds. Recently, we have
analyzed\cite{Svit} the reorientational motion in ferroelectric KTa$_{1-x}$Nb$_{x}$O$%
_{3}$ crystals. In this paper we discuss the role of the
relaxational and reorientational dynamics in the scattering processes from
PMN crystal.

In the paper presented we offer a complex analysis of temperature
dependencies of polarized (VV) and depolarized (VH) Raman spectra from a
single crystalline PMN sample. This analysis is based on multiple peak
decomposition. An attempt to apply similar kind of analysis has
earlier been made by Siny et al.\cite{siny-cp}. But, it was restricted to the low frequency VV spectra in a limited temperature range (with maximum temperature $\sim $620~K), concentrating on the central
peak (CP) only. To obtain the complete picture, we consider the temperature
evolution of the whole Raman scattering spectrum (up to 1000~cm$^{-1}$) in
the 100-1000~K temperature range.\ We interpret our results in connection
with recently published neutron scattering data and with the concept of
interaction of phonons with internal relaxational motion. As a result of
this work, we present our view on the picture of structural transformations
in PMN crystal and on the origin of its Raman spectrum.

\begin{figure}[th]
\includegraphics[width=1.1\columnwidth]{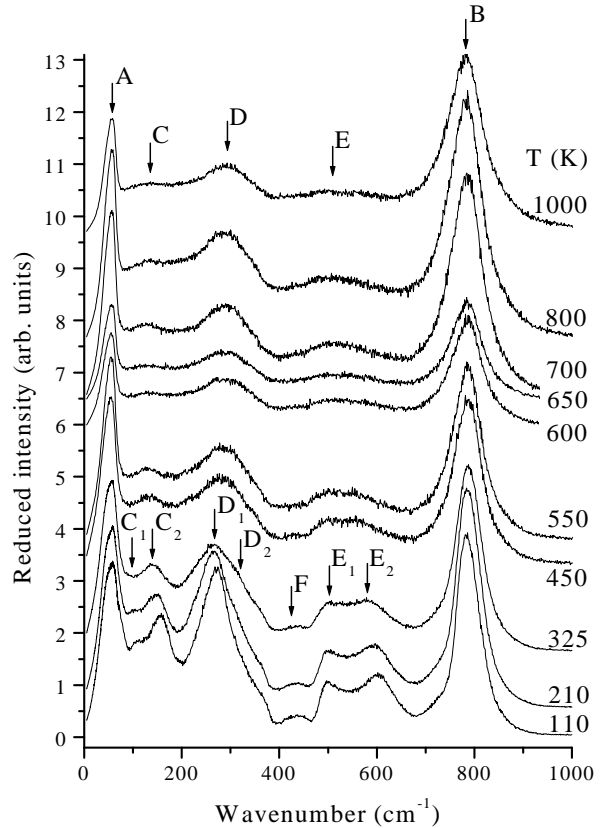}
\caption{Examples of Raman scattering spectra measured at various
temperatures without polarization analysis and corrected by the population
factor. }
\label{cascade}
\end{figure}

\begin{figure*}[t]
\includegraphics[width=2\columnwidth]{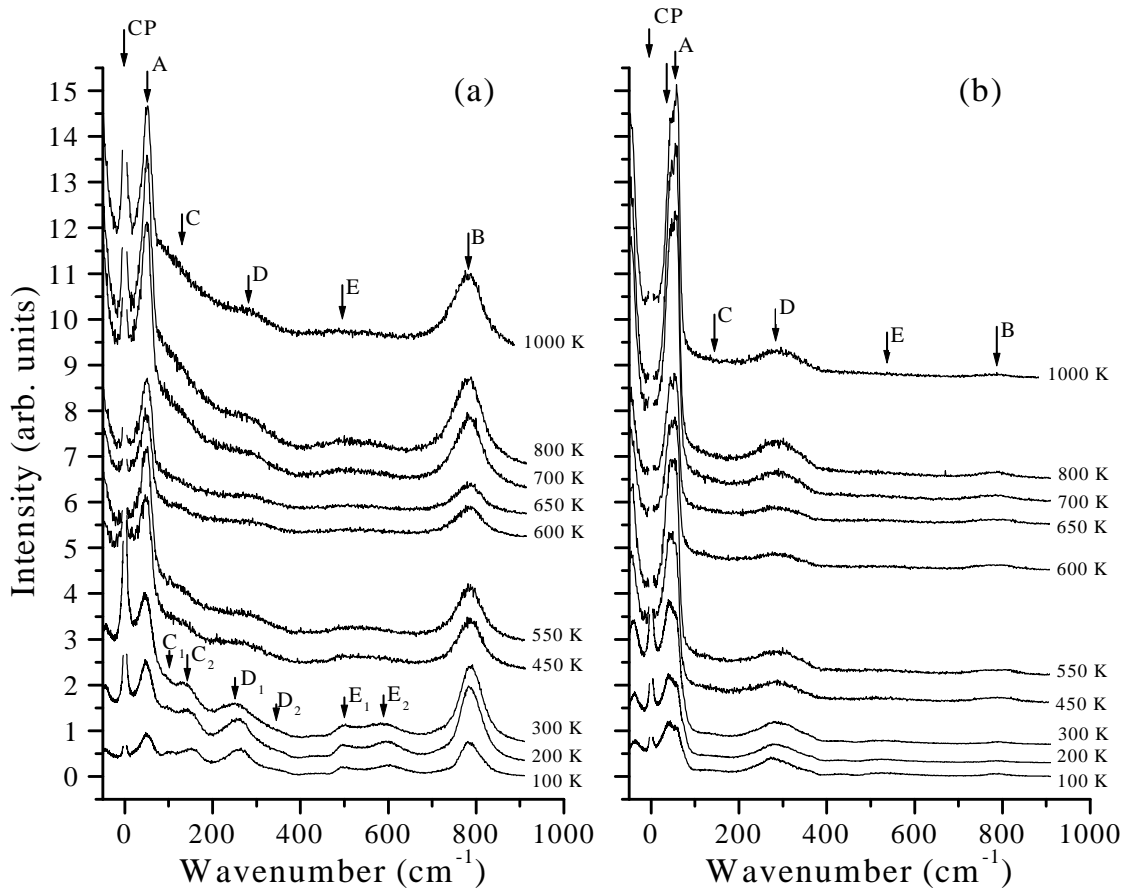}
\caption{Examples of Raman scatteting spectra measured in VV (a) and VH (b) geometries upon cooling from 1000 to 100 K.}
\label{vv-vh}
\end{figure*}

\section{Experimental setup and results}

We have investigated the Raman scattering from $\left\langle
100\right\rangle $ cut PMN\ single crystal. The crystal was grown by the
high temperature solution technique using 30~wt\% PbO as flux. The growth
conditions were optimized based on the pseudo-binary phase diagram
established for PMN and PbO\cite{ye-1}. The as-grown crystal exhibits a
pseudo-cubic morphology with $\left\langle 100\right\rangle $ cub growth
steps. A $\left\langle 100\right\rangle /\left\langle 110\right\rangle $
cub-oriented crystal of a volume of 53~mm$^{3}$ was cut from a large as-grown
crystal and polished with fine diamond paste (down to 0.25~mm). It showed
very high optical quality and satisfied the requirements for light scattering
studies. The scattering was excited by propagating in $\left\langle
100\right\rangle $ direction 514.5~nm light from a 200~mW Ar$^{+}$-ion
laser, focused to a 0.1~mm spot. The scattered light was collected at an
angle of 90$^{\circ }$ with respect to the incident beam (i.e., in $%
\left\langle 010\right\rangle $ direction) by a double-grating ISA Jobin
Yvon spectrometer equipped with a Hamamatsu photomultiplier R-649. For most
of the measurements, the slits were opened to 1.7~cm$^{-1}$. However, in
order to acquire more precise data in the central peak region, at the
temperatures close to the maximum of the dielectric constant ( $100<T<350$%
~K), the slits were narrowed to 0.5~cm$^{-1}$. Each polarization of the
scattered light, $\left\langle x|zz|y\right\rangle $ (VV) and $\left\langle
x|zx|y\right\rangle $ (VH), was measured separately. In order to exclude
differences in sensitivity of the monochromator to different polarizations
of the light, a circular polarizer was used in front of the entrance slit.
For control purposes, we also took measurements without polarization
analysis. Finally, to protect the photomultiplier from the strong Raleigh
scattering, the spectral region from -4 to +4~cm$^{-1}$ was excluded from the
scans. The data were collected in the temperature range from 1000 to 100~K.
The cooling rate was 0.5-1~K/min. Every 50-20~K the temperature was
stabilized and the spectrum recorded. We agree with Ref.\cite{ohwa} that
the reproducibility of the spectra in the temperature range approximately
between 200 and 350~K was not very good. In this range, the spectra exhibited
a strong dependence on the cooling rate and other experimental conditions.

Fig.\ref{cascade} presents examples of light scattering spectra measured at
different temperatures without polarization analysis. To facilitate comparison, we corrected the spectra by the Bose population factor:

\begin{equation}
F(f,T)=\left\{ 
\begin{array}{c}
n(f)+1,\text{ for Stokes part} \\ 
n(f),\text{ for anti-Stokes part}%
\end{array}%
\text{ }\right. \text{, }  \label{boze}
\end{equation}%
where 
\begin{equation*}
n(f)=(exp(hf/kT)-1)^{-1}.
\end{equation*}

\noindent Figure \ref{vv-vh} demonstrates scattering spectra measured in VV (a) and VH (b) geometries at different temperatures. These spectra we present in uncorrected, "as-measured", form. 

\begin{figure*}[t]
\includegraphics[width=1.8\columnwidth]{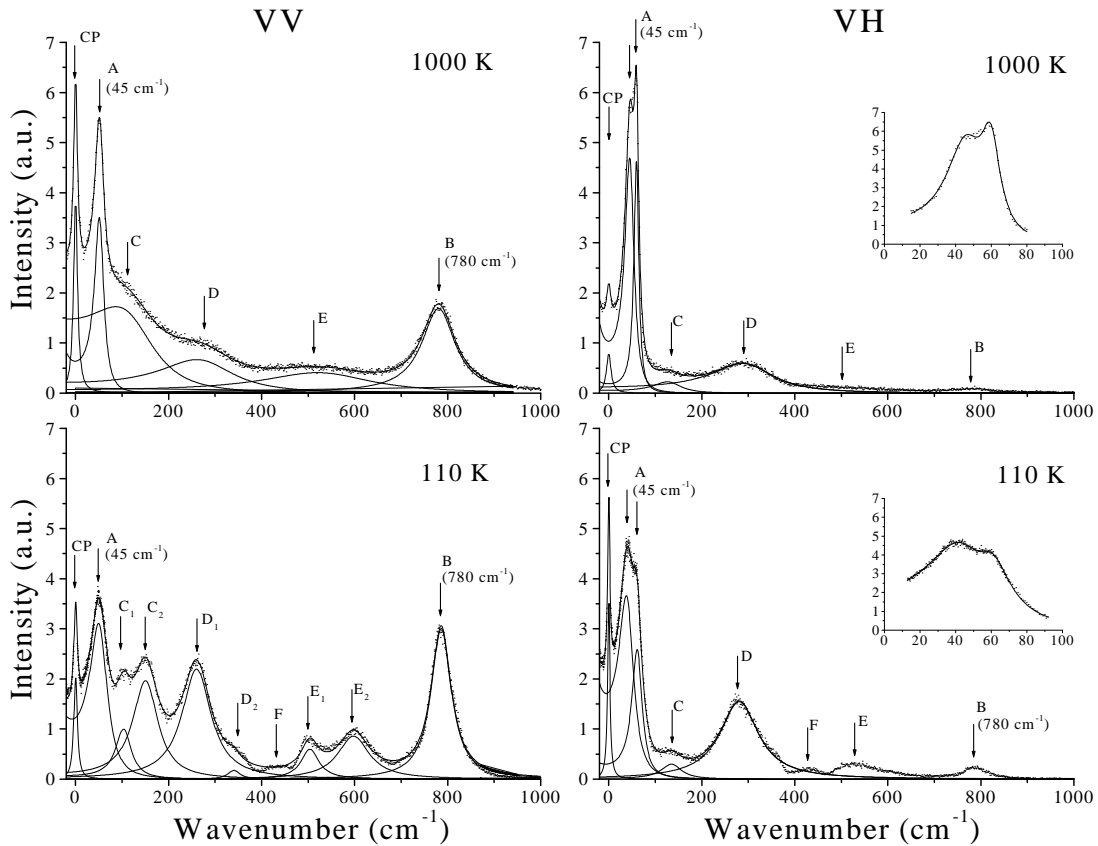}
\caption{Examples of multiple peak decomposition of Raman spectra
measured at 1000 and 110~K both, in VV and VH geometry. For convenience,
major peaks are labeled with letters. Inserts show examples of fits
for the line A using the model of the two coupled harmonic oscillators, as
described in Discussion.}
\label{FitExamples}
\end{figure*}

As can be seen from Figs. \ref{cascade} and \ref{vv-vh}, our spectra are consistent with those
from Refs.\cite{husson} and \cite{ohwa}. In the
high temperature region, a typical spectrum consists of two strong lines
centered approximately at 45~cm$^{-1}$ and 780~cm$^{-1}$ (labelled A and B)
and of three broad bands (C, D, E). Line A exhibits fine structure, which is
explicitly seen on the spectra measured with polarization analysis (see Figs. \ref{vv-vh} and
\ref{FitExamples}). Lowering temperature leads to the
splitting of the broad bands C, D and E into a number of narrower lines
(labelled by indices) and to appearance of new line F. 

As seen from Fig. \ref{cascade}, starting
from 1000~K, the reduced intensity of the scattering, first grows until the temperature
of $\sim700$~K then sharply decreases to a minimum located close to the Burns temperature T$_{d}\approx620$~K, after which it increases again. At $\sim550$~K its strength gets restored. Below this temperature, the intensities of the major lines (especially A and B) are determined primarely by the Bose factor. I.e. being corrected by it, they do not show significant temperature changes, demonstrating characteristic first-order behavior. The shape similarity of the low- and high- temperature spectra shows that even at 1000~K the scattering has first-order character. In the perfecly cubic crystal, first-order scattering is prohibited. Therefore, one should assume the presence of distortions in the form of lower symmetry clusters, like Fm3m or R3m. At high temperature, these clusters might be present in the dynamic form, having short lifetime. Lowering the temperature, their lifetime increases and they become progressively more static. The observed around T$_{d}$ decrease of Raman intensity is also clearly seen in Fig. \ref{vv-vh}. However, there are no reports on this effect in the literature; it might have been overlooked. This intensity drop indicates worsening of the optical quality of the crystal. In our opinion, it is consistent with the formation of Fm3m clusters in their evolution from dynamic to static form. The splitting of peaks is associated with the development of R3m clusters. The first signs of the splitting appear at T$_{d}$. Strarting from T$_{f}\sim350$~K (which is also close to where the dispersion of the dielectric constant appears), this effect becomes even more definite and the intensities of the components begin to grow. Below T$_{do}\sim210$~K, which is the temperature of electric field-induced ferroelectric phase transition\cite{Ye} , the shape of the spectra is relatively stable. 

A comprehensive analysis of our data and their comparison with results of other experiments allows to understand the meaning of the mentioned above temperatures T$_{d}\approx620$~K, T$_{f}\approx350$~K and T$_{do}\approx210$~K in the temperature evolution of the Raman spectra and the lattice structure. 

\section{Multiple-peak decomposition of spectra procedure.}

To be able to analyze the data, we decomposed the measured spectra using
multiple peak fitting procedure. It was found that satisfactory fit could be
achieved with the assumption that the central peak has a Lorentzian shape and that each of the other peaks is described by the spectral response
function, i.e. damped harmonic oscillator, modified by the population factor (%
\ref{boze}):

\begin{equation}
\Phi _{i}\sim \frac{\Gamma _{i}f_{0i}^{2}f}{(f^{2}-f_{0i}^{2})^{2}+\Gamma
_{i}^{2}f_{0i}^{2}}F(f,T)\text{ ,}
\end{equation}

\noindent where $\Gamma _{i}$ and $f_{0i}$ are the damping constant and the
mode frequency.

As all of the peaks are much better resolved at low temperatures, we started
our fitting procedure at the low-temperature end of the data set (at 110~K)
and, then, observed the evolution of the peaks with increasing
temperature (i.e., in order, opposite to the order of measuring). At
the same time, we tried to minimize the number of peaks necessary to achieve
a reasonably good fit. The control data set (measured without polarization
analysis) has been used to calibrate the positions and widths of the weak and poorely resolved
peaks from the VV and VH data sets. Since a large number of parameters is
involved, results of a particular fit may depend on their initial values. To
stabilize the results, the best-fit values of parameters obtained at the
previous temperature were used as initial values for the next one. In this
manner, several sets of fits were obtained and analyzed. It is remarkable,
that in all of them the major parameters showed the same trends of
behavior, confirming the importance of the mentioned above temperatures: T$_{d}\approx620\pm 50$~K, T$_{f}\approx350\pm 25$~K and T$_{do}\approx210\pm 25$~K. These trends are summarized in Table \ref{peakproperties}. One of the sets have been selected to report the most interesting results in detail. Examples of the fits at the temperatures of 1000 and 110~K in VV and VH geometries are shown in Fig. \ref{FitExamples}. For clarity, we describe the observed phenomena from high to low temperatures, following the same order as in measurements (unless the opposite stated).

\begin{table*}[th]
\caption{Major peaks and their response to the temperature restrictions on ion motion. 
Only changing properties are shown. T$_{d}\approx620$~K is also characterized by the total loss of
intensity of scattered light.}
\label{peakproperties}%
\begin{ruledtabular}
\begin{tabular}{ccllll}
Peak   & Polari-& $Region$ $I.$ Unrestricted  & $Region$ $II.$ Restricted                         & $Region$ $III.$ Formation of   \\ 
       & zation & dynamical clusters          & dynamical clusters                                & static R3m clusters            \\ 
       &        & T$\gtrsim$(T$_{d}$=620~K)   & (T$_{d}$=620 K)$\gtrsim$T$\gtrsim$(T$_{pr}$=350~K)& (T$_{pr}$=350~K)$\gtrsim$T      \\
\hline
\hline
Central& VV    & Intensity decreases          & Intensity increases                     & Intensity has a maximum at 300 K  \\
       &       & broadens                     & narrows                                 &            \\
       & VH    & Intensity decreases          & Intensity increases                     & Intensity has a maximum at 300 K \\
       &       &                              & width has a maximum at 450 K            &            \\
\hline
Peak A & VV    &                              & Softens, broadens                       &Hardens and narrows \\
       & VH-I  & Broadens                     & Broadens                                & Broadens         \\
       & VH-II & Softens                      & Hardens, intensity has minimum at 400 K & Broadens and grows \\
\hline
Peak C & VV    & Hardens and narrows          & Fine structure appears                  & Splits in two \\
       & VH    & Softens                      & Hardens and narrows                     & Narrows  \\
\hline
Peak D & VV    &                              & Splits in two                           &                                 \\
       & VH    & Softens                      & Hardens                                 & Softens and narrows \\
\hline
Peak F & VV    &                              & VV component appears                    & VH component appears \\
\hline
Peak E & VV    &                              & Fine structure appears                  & Splits in two, VH appears \\
\hline
Peak B & VV    & Hardening                    & Hardening                               & VH appears \\
\end{tabular}
\end{ruledtabular}
\end{table*}

\begin{figure}[ht]
\includegraphics[width=.9\columnwidth]{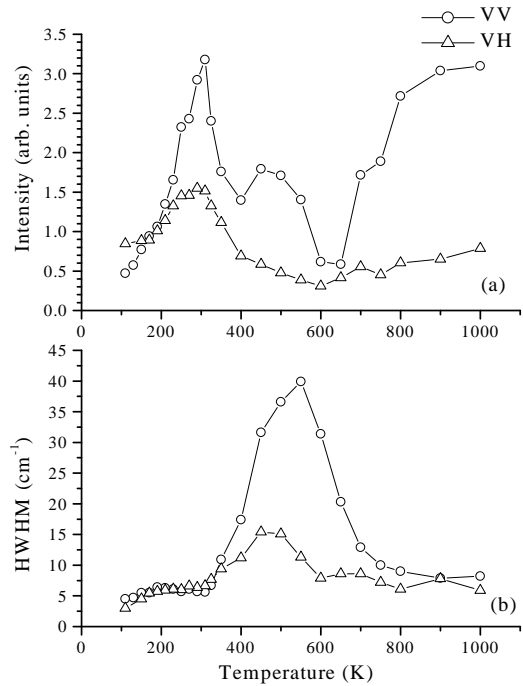}
\caption{Temperature dependencies of the intensity (a) and half-width (b) of
the Lorentzian approximation for the central peak in VV (circles) and VH
(triangles) geometries of experiment.}
\label{central}
\end{figure}

\section{Most interesting results of decomposition and their analysis.}

\subsection{Central peak.}

The temperature dependences of the fitting parameters for the central peak
(CP) are presented in Fig. \ref{central}. Circles correspond to the VV and
triangles to the VH component of the peak. The existence of the CP is a direct consequence of the lattice relaxations, which are very sensitive to the restrictions imposed by the low-symmetry clusters. If relaxations are fast\cite{MichelA}, the CP is low-intense and broad, whereas their slowing causes growth and narrowing of the peak. 

We would like to point out a striking similarity of the temperature behavior of the CP in PbMg$_{1/3}$Nb$_{2/3}$O$_{3}$ (Fig. \ref{central}) and in KTa$_{0.85}$Nb$_{0.15}$O$_{3}$ (Fig.3 in Ref. \cite{Svit}) crystals. We have shown\cite{Svit} that the temperature behavior of the CP in KTN can be explained by the model involving the relaxational motion of off-centered Nb ions and its progressive restriction with temperature decrease. In the cubic phase, Nb ions are allowed to reorient amongst eight equivalent $<$111$>$ directions. The appearance of the PNR's followed by a sequence of phase transitions down to a rhombohedral R3m phase, limits the ion motion to four, two and, finally, locks it in only one site. This model is in agreement with the diffuse neutron scattering studies of similar system\cite{grace}. The similarity of the CP behavior in these two very different materials, suggests that the temperature evolution of the polar clusters in PbMg$_{1/3}$Nb$_{2/3}$O$_{3}$ passes through the similar stages as in KTa$_{0.85}$Nb$_{0.15}$O$_{3}$ (KTN), however, it is not accompanied by appearance of the
long-range order. 

Starting from the high-temperature end (Fig. \ref{central}), the first important feature is the strong and narrow scattering in the VV geometry accompanied by a relatively weak scattering in the VH geometry. This indicates the presence of a symmetric slow relaxational motion, involving 180$^\mathrm{o}$ reorientations of ions. Lowering the temperature, starting from $\sim900$~K, a cessation of this motion causes an intensity decrease of the CP. Slow at first, this decrease becomes sharper and reaches minimum near T$_{d}\approx620$~K. The prohibition of 180$\mathrm{o}$ reorientations introduces anisotropy into the lattice causing the distinguishability between Mg and Nb occupied cites and to formation in the 1:1 ordered areas of the superstructure with, on average, Fm3m symmetry. It also imposes first restrictions on the reorientational motion of the dynamical R3m polar nanoregions. Now, they can reorient only amongst four neighboring $<$111$>$ directions, forming, on average, tetragonal-like distortions. These processes are accompanied by the appearance of large (of the order of wavelength of light) dynamic fluctuations causing worsening of the optical quality of the sample.

With further decrease of the temperature, the optical quality of the crystal improves. From $\sim550$~K, the four-cite reorientational motion of the PNR's starts to slow down, which is marked by the narrowing of the VV component of the CP and increase of its VV and VH intensities. The simultaneous broadening of the VH component up to the maximum at $\sim450$~K, indicates rearrangements in the crystalline structure leading to the appearance of the new type of restrictions on the ion motion. Analogy with KTN suggests that below $\sim450$~K the motion of R3m clusters becomes limited to two neigboring $<$111$>$ orientations, averaging in monoclinic-like distortions. Such a rearrangement causes some decrease in intensity of the VV component (with minimum at $\sim400$~K), while the VH intensity keeps growing. Below $\sim400$~K, the slowing down of the two-site relaxational motion causes narrowing the CP and increase of intensities of both components. From T$_{f}\approx350$~K, these effects become especially dramatic. Further temperature decrease, starting from $\sim300$~K, leads to the complete prohibition of intersite reorientational motion of R3m clusters, i.e. to appearance of static R3m clusters. This is marked by a sharp decrease in intensity of both components of the CP. At the temperature T$_{do}\approx210$~K, the process of conversion of PNR's from dynamic to static form (freezing) is primarily finished. Below this temperature, the central peak is narrow and its intensity is small in both scattering geometries.

We should mention that the result of our analysis of the VV component of the CP exhibits
similar tendencies to those in Ref.\cite{siny-cp}, except at the low
temperature end. The discrepancy occurs due to the difference in
experimental technique. In the central peak region, we did the measurements
with much higher resolution (0.5~cm$^{-1}$, as
compared to 2~cm$^{-1}$ from Ref.\cite{siny-cp}). Consequently, we were able to provide a more correct separation of the slow relaxations from the elastic scattering, which was especially
important at low temperatures.   

\begin{figure}[th]
\includegraphics[width=.9\columnwidth]{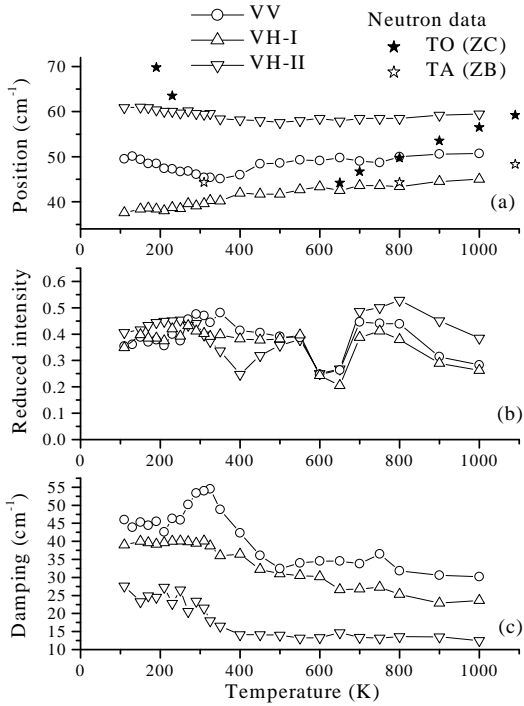}
\caption{Temperature dependences of the fitting parameters: position (a),
reduced intensity (b) and damping constant (c) for the triplet line A
(located at around 45~cm$^{-1}$). Phonon frequencies, measured by neutron
spectroscopy\protect\cite{GehWak, WakStock, GehVak, NaberVak}, are shown for
comparizon.}
\label{line45}
\end{figure}

\subsection{Line A.}

Figure \ref{line45} presents the temperature evolution of the
fitting parameters of the peak A, showing its position (a), reduced intensity (b)
and damping constant (c). This peak has a triplet structure, containing one component in VV
(circles) and two components in VH (up and down triangles) geometries. The fitting parameters for this peak exhibit changes at mentioned above temperatures T$_{d}$, T$_{f}$ and T$_{do}$, confirming their importance for the structural evolution of the crystal. However, the origin of this peak (see Table. \ref{RamanConc}) requires qlarification. From a comparison with the frequencies of the phonon modes determined from neutron scattering, it is clear that this line cannot be due to the zone center soft TO$_{1}$ mode (black stars in Fig. \ref{line45}). On the other hand, the lower frequency VH component, and possibly the VV component could be due to the disorder-induced zone boundary scattering on TA phonon (white stars in Fig. \ref{line45}). However, the higher frequency VH component would still not be accounted for.

In an attempt to account for both VH components simultaneously, we have tried to make use of a coupled oscillator model. In the approximation of linear coupling between two harmonic oscillators\cite%
{currat} this approach gains:

\begin{equation*}
I(\omega )=[n(\omega )+1]\left( \frac{CY-BZ}{B^{2}+C^{2}}\right) ,
\end{equation*}

\noindent where

\bigskip

\noindent $B=(\omega _{1}^{2}-\omega ^{2})(\omega _{2}^{2}-\omega
^{2})+\omega ^{2}(\gamma _{12}^{2}-\gamma _{1}\gamma _{2})-\omega _{1}\omega
_{2}\Delta _{12}^{2},$

\noindent $C=\omega \lbrack \gamma _{1}(\omega _{2}^{2}-\omega ^{2})+\gamma
_{2}(\omega _{1}^{2}-\omega ^{2})-2(\omega _{1}\omega _{2})^{1/2}\Delta
_{12}\gamma _{12}],$

\noindent $Y=S_{1}^{2}(\omega _{2}^{2}-\omega ^{2})+S_{2}^{2}(\omega
_{1}^{2}-\omega ^{2})-2S_{1}S_{2}(\omega _{1}\omega _{2})^{1/2}\Delta _{12},$

\noindent $Z=\omega (S_{1}^{2}\gamma _{2}+S_{2}^{2}\gamma
_{1}-2S_{1}S_{2}\gamma _{12}).$

\bigskip

\noindent In these equations $\omega _{1}$ and $\omega _{2}$ are the
resonant frequencies, $\gamma _{1}$ and $\gamma _{2}$ are the damping
constants, $S_{1}$ and $S_{2}$ are the structure factors for two
oscillators, $\Delta _{12}$ and $\gamma _{12}$ are the real and imaginary
parts of the coupling constant.

\begin{figure}[th]
\includegraphics[width=.9\columnwidth]{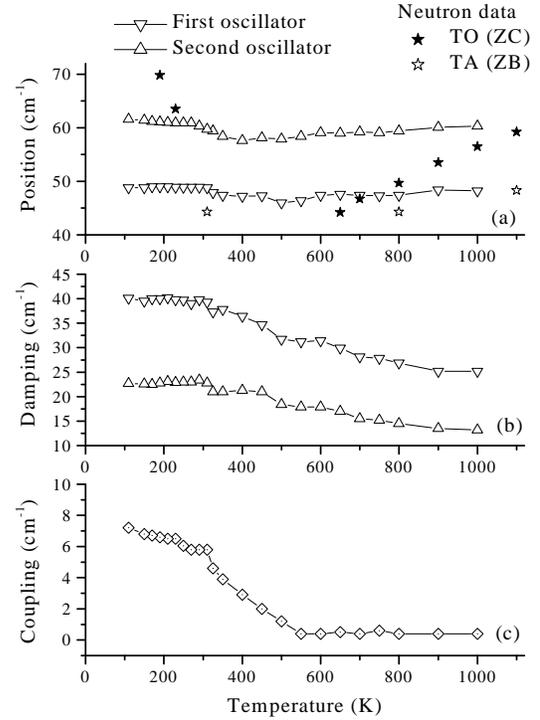}
\caption{Temperature dependences of the fitted parameterst of the line A VH
component (located at 45 cm$^{-1}$) using the model of coupled
harmonic oscillators. From top to bottom: resonant frequencies (a), damping
constants (b) and coupling constant (c). Phonon frequencies, measured by
neutron spectroscopy\protect\cite{GehWak, WakStock, GehVak, NaberVak}, are
shown for comparizon.}
\label{coupled}
\end{figure}

It was shown\cite{barker}, that a unitary transformation can be found, which
would set either real or imaginary part of the coupling to be equal to zero.
We have chosen to use imaginary coupling and set $\Delta _{12}=0$. We also set the structure factors $S_{1}$ and $S_{2}$ to
be the same at all temperatures. The rest of the parameters
were treated as being temperature dependent. In the result of fitting, we
have found that the values of structure factors can be approximated as $%
S_{1}=22.3$~cm$^{-1}$ and $S_{2}=20.6$~cm$^{-1}$. The other parameters as a function of temperature are shown in Fig.\ref{coupled}: the best-fit frequencies (a), the damping constants (b) and the coupling coefficient (c).

The examples of the fits (inserts in Fig.\ref{FitExamples}) demonstrate that
this model gives a good approximation to the shape of the peak A. However, a comparizon of the best-fit frequencies (Fig. \ref{coupled}a) with the neutron scattering data\cite{GehWak, WakStock, GehVak, NaberVak} rules out the possibility to explain the higher-frequency VH component by the interaction between ZC TO$_{1}$ (black stars) and ZB TA (white stars) modes. The lower frequency oscillator can still be associated with disorder-induced scattering on ZB TA mode.

The possible explanation of the presence of two VH components in the peak A can be in the correspondence to TA modes with different polarizations in different crystalline planes of a distorted lattice. (A similar effect we have seen in monodomain KTa$_{1-x}$Li$_{x}$O$_{3}$, where one or the other TA polarization was observed, depending on the orientation of the electric field.) These differently polarized modes are influenced by the polar nanoregions, which explains the increase of damping (Fig. \ref{coupled}b) and coupling (Fig. \ref{coupled}c) coefficients with lowering temperature.

It is also necessary to point out an observation concerning the temperature dependences of the frequencies of the components of the triplet A in approximation of independent oscillators (Fig. \ref{line45}a). These dependences seem to show anomalies at intersections with ZC TO$_{1}$ modes (black stars). For example, the VV component exhibits deeps at $\sim700$ and 350~K and VH-I component has a knee at $\sim400$~K.

\begin{figure}[th]
\includegraphics[width=.9\columnwidth]{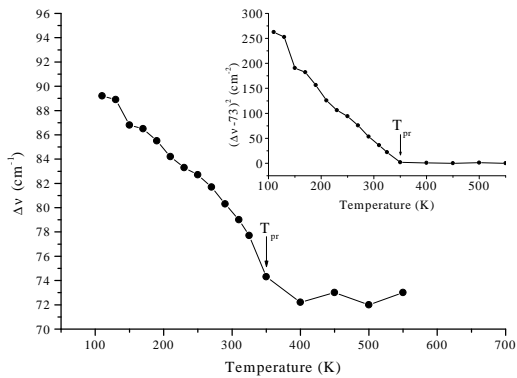}
\caption{Amplitude of splitting between the components of the line E. The insert demonstrates that this splitting follows the critical dependence.}
\label{split}
\end{figure}

\subsection{The most important parameters of other lines.}

Among the changes occuring to the Raman spectrum with cooling the sample, the splitting of the phonon lines desrves special attention, since such a splitting is a direct indicator of the modifications of the local structure. As we already mentioned, the splitting occurs to the lines C, D, and E, which is especially well seen in the VV geometry (Fig. \ref{vv-vh}a). At high temperatures, their widths are very big, so the lines almost merge in a single broad shoulder. Lowering the temperature, they become more discernible and the fine structure gradually develops. The first signs of it appear at T$_{d}\approx620$~K. Below T$_{pr}\approx350$~K, the fine structure becomes more obvious. These observations support the expressed above idea that the scattering originates from the distorted regions. Highly dynamic and disordered at high temperature, they become progressively more static with cooling down, and their motion becomes more correlated. This process is reflected by the temperature evolution of the lines C, D, and E.

Below the temperature of T$_{pr}\approx350$~K, the splitting value of the line E exhibits especially interesting trend. As seen from Fig. \ref{split}, with temperature decrease below T$_{pr}$ the distance between the E-line components increases. Moreover, this increase exhibits the characteristic order parameter behavior, following the (T$_{pr}-$T)$^{-1/2}$ dependence, which is demonstrated in the insert to Fig. \ref{split}. It also exhibits similar temperature tendency to that described in Ref. \cite{pros1}. Detailed understanding of this phenomenon requires further work, and it is a subject of separate paper.

\begin{figure}[th]
\includegraphics[width=.9\columnwidth]{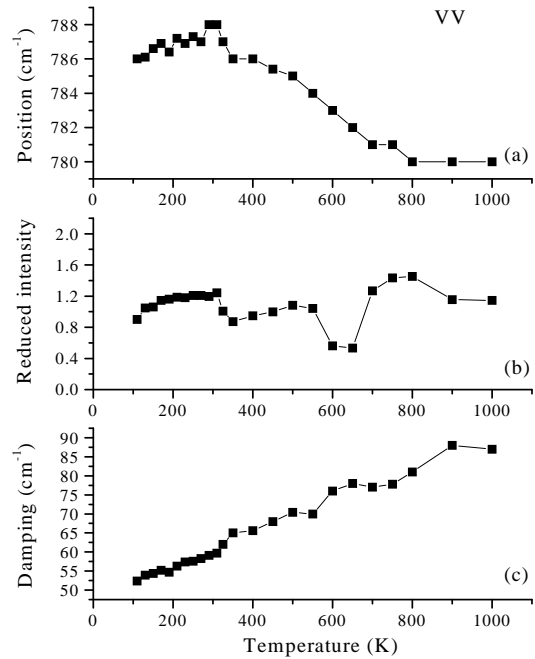}
\caption{Temperature dependencies of the fitting parameters: position (a),
reduced intensity (b) and damping constant (c) of the line B (located at $%
\thicksim $780~cm$^{-1}$.)}
\label{line780}
\end{figure}

Finally, we would like to stop on the line B located at the upper end of the spectrum. Fig.~\ref{line780} demonstrates the fitting paramenters for this line: position (a), reduced intensity (b), and damping coefficient (c). This line is very strong in the VV geometry but it is rather weak in the VH one (i.e. characterized by A$_{1g}$ symmetry). It seems to be the least influenced by the ordering processes in the crystal. However, the temperature evolution of its parameters reflects development of the low-temperature phase, confirming the expressed above ideas about formation of dynamic polar clusters at T$_{d}\approx620$~K, their slowing down, and appearance of the static order at T$_{pr}\approx350$~K. As it is seen from the Table~\ref{RamanConc}, the appearance of this line in the Raman spectrum has several contradictory explanations. However, according to the recent first-principle calculations\cite{pros2}, this mode most likely is a consequence of oscillations of oxygen ions. 

\section{Conclusion}

This paper offers a comprehensive analysis of the Raman scattering spectra of PbMg$_{1/3}$Nb$_{2/3}$O$_{3}$ in the temperature range between 100 and 1000~K. From our analysis, the structural evolution of the lattice occurs in several stages. Even at 1000~K, the PMN crystal is characterized by the presence of dynamic distortions of the lattice from cubic symmetry, which are responsible for existence of first-order Raman scattering at such a high temperature. At the Burns temperature T$_{d}\approx620$~K, the first restrictions on the reorientational motion of the off-centered ions appear. These restrictions lead to slowing of the motion and growth of correlations between neighboring regions with R3m symmetry. They also introduce distinguishability between sites occupied by Mg and Nb, which causes onset of Fm3m symmetry in the 1:1 ordered areas. Further cooling leads to progressive growth of restrictions on ionic motion and, starting from T$_{pr}\approx350$~K, to the appearance of the static polar nanoregions. At T$_{do}\approx210$~K, the formation of the low-temperature phase is primarily finished. The shape of the phonon lines is determined by multiphonon processes involving phonons with different polarizations propagating in different directions, which interact with dynamic and static disorder.

\section{Acknowledgement}

Authors are very grateful to D.La-Orauttapong and G.Yong for useful advice and helpful discussions. This work
was partially supported by the ONR grant \#N00014-99-1-0738.

\end{document}